\documentclass[12pt]{iopart}

\usepackage{graphicx}
\usepackage{epsfig}
\usepackage{placeins}

\begin{document}

\title{Statistics of acoustic emission in paper fracture: precursors and criticality }
\author{J. Rosti, J. Koivisto, M.J. Alava}
\address{Department of Applied Physics, Helsinki University of
Technology, FIN-02015 HUT, Finland}
\ead{jro@fyslab.hut.fi}

\maketitle
\pacs{62.20.M-,05.40.-a, 46.50+a}

\begin{abstract}

We present statistical analysis of acoustic emission (AE) data from
tensile experiments on paper sheets, loading mode I, with samples
broken under strain control. The results are based on 100
experiments on unnotched samples and 70 samples with a long initial
edge notch. First, AE energy release and AE event rates are
considered for both cases, to test for the presence of "critical
points" in fracture. For AE energy, no clear signatures are found,
whereas the main finding is that the event rate diverges when a
sample-dependent "critical time" of the maximum event rate is
approached. This takes place after the maximum stress is reached.
The results are compared with statistical fracture models of
heterogenous materials. We also discuss the dependence of the AE
energy and event interval distributions on average event rates.

\end{abstract}

\maketitle

\date{\today}

\section{Introduction}

Fracture of heterogenous material exhibits scaling properties familiar
from statistical physics. Scaling laws of energy release and fluctuations in temporal
statistics tempts one to interpret fracture via concepts of
criticality and phase transitions. However, simple models have
failed to convey the phenomenology and the origins of
observed scaling laws are lacking. In this work we
have performed extensive set of tensile experiments on paper sheets in order to
discuss the concept of criticality in  mode I loading with imposed strain. \cite{petri1, maes}.

Typical stress-strain curve in tensile fracture of paper samples
exhibits some non-linearity before the maximum stress, $\sigma_c$.
This is a result of both plastic, irreversible deformation and to a
much lesser degree of loss of elastic stiffness, or damage
accumulation \cite{AlavaNiskanen}. We study the damage accumulation
by using ``acoustic emission'' (AE). This is the release of elastic
energy due to microcracking on various scales ranging from far below
the fiber size to, perhaps, millimeters as in the individual
advances made by a notch in a tensile test. AE studies have been
done in a wide variety for materials in science and engineering
\cite{AE}. The common features in most are that the material failure
process is complicated and disorder is present in the
structure. Numerous studies have elucidated the statistical laws
that describe AE. In general, these relate to usual mode I or mode
III -type loading conditions, and the common feature is that they
exhibit{\em scalefree} features. The probability distribution
function (pdf) of event energies follows most often a power-law,
with exponents in the range $\beta = 1\dots 2$. The event
intervals are, similarly, found to obey such fat-tailed
pdf's. \cite{sethna, laurinprl, advphys, maes, jphys}

%%% MOTIVATION
Here we consider paper as a test material or case of critical
divergences during the fracture in the presence of structural
randomness. This idea of criticality can be  formulated in various
ways, but essentially it implies the presence of a finite time
singularity, so that a quantity such as the AE energy release during
the approach of the failure becomes a function of $1/(t_c - t)$ where $t_c$ is the
lifetime. In the following, we focus on the non-stationary AE
signal from tensile experiments with imposed strain and study energy
and event statistics of AE near the failure. We study in detail
local averages of e.g released acoustic emission energy when
$t_c$, or critical point, is approached. The relation between critical
point $t_c$ and the time at the maximum stress $\sigma_c$ is discussed. 

The total number of observed acoustic emission events
in all experiments is 800921. The high number of events in a single 
experiment (few thousands of events, comparable to rock and concrete fracture experiment cases  
\cite{finck,tham}) combined with the
large number of samples allows us to
study the non-stationarity of the acoustic emission signal. 

What kind of
AE signature is produced by driving the fracture, e.g. by imposing a
constant strain or creep stress, is not well understood. Thus, understanding the statistical
properties of the acoustic emision and fracture precursors could provide a way to distinguish between
different modes of loading. 
One possible application of this is in creating analogies with
experimental findings in seismicity \cite{rosti}.

The main result of the analysis is that we can distinguish two
fundamental cases: the dynamics in samples without a dominating
notch and dynamics in the presence of a large defect with slow, stable crack
growth. In the former case, there is evidence of such a divergence,
upon approach of a sample-dependent critical time. This takes place
after the sample maximum stress is reached. A qualitative
explanation is perhaps offered by the dynamics which stems from the
nucleation and growth of a dominating microcrack.

To discuss the divergence, we consider simple models of statistical
fracture. We show that variants of a minimal quasi-static fracture
(random fuse network, RFN) model do not exhibit such features as the
data. It is well-known that the RFN and other models do not reproduce
quantitatively e.g. the exponents that characterize AE statistical
properties. It also turns out that the divergence is not found in
such models, however a RFN variant with gradual failure gets
slightly closer \cite{li}.

The structure of the rest of this paper is as follows. In Section 2,
we discuss the background and details of the experiments. Section 3
presents various aspects of the data analysis, and is finished with
a subsection on modeling attempts. Section 4 presents discussion
and summary.

\section{Experiments}

Normal copy paper samples were tested in the cross direction
on a mode I laboratory testing machine of type MTS 400/M.
The deformation rate $\dot{\epsilon}$ was 10~mm min$^-1$.
The AE system consisted of a piezoelectric receiver, a rectifying
amplifier and continuous data-acquisition. The data-acquisition was free of deadtime.
The stress was measured simultaneously directly from MTS 400/M using same
time-resolution as AE-signal; thus the accuracy of time synchronization of the
acoustic emission and stress-strain curves was below 10~$\mu$s.

During the experiment, we acquired bi-polar acoustic amplitudes
simultaneously on two channels by piezocrystal sensors,
as a function of time. Two transducers
were attached directly to paper without a coupling agent.  Each channel 
has 12-bit resolution and
a sampling rate of 312~000~Hz. The transmission time from event origin to
sensors is in the order of 5~$\mu$s. The acoustic channels were first
amplified and after that held using sample-and-hold circuit.

70 samples had initial notches of size 10~mm
to achieve stable crack growth and 100 samples were intact.
Sample geometry was 150~$\times$~50~mm$^2$. Samples were tested in standard conditions: temperature of
22 degree in Celcius and relative humidity of 30\%.

The acoustic time-series was post-processed after the measurement by
detection of continuous and coherent events, and the calculation of
an event energy $E_i$ is done as the integral of squared amplitudes
within the event: $E_i= \int A^2(t') dt'$. The event arrival time
$t_i$ was taken from an instant when the amplitude raises above the 
threshold level. Dynamic range of the measurement device was 54~dB.
The  ensuing discrete set of events is characterized by set of pairs:
$\{(t_1, E_1), (t_2, E_2), \dots\}$. Simultaneously, we measured the
stress signal $\sigma(t)$ and, on the basis of that, we can define quantities
also as a function of stress $\sigma(t)$.

\begin{figure}
\centering
\epsfig{figure=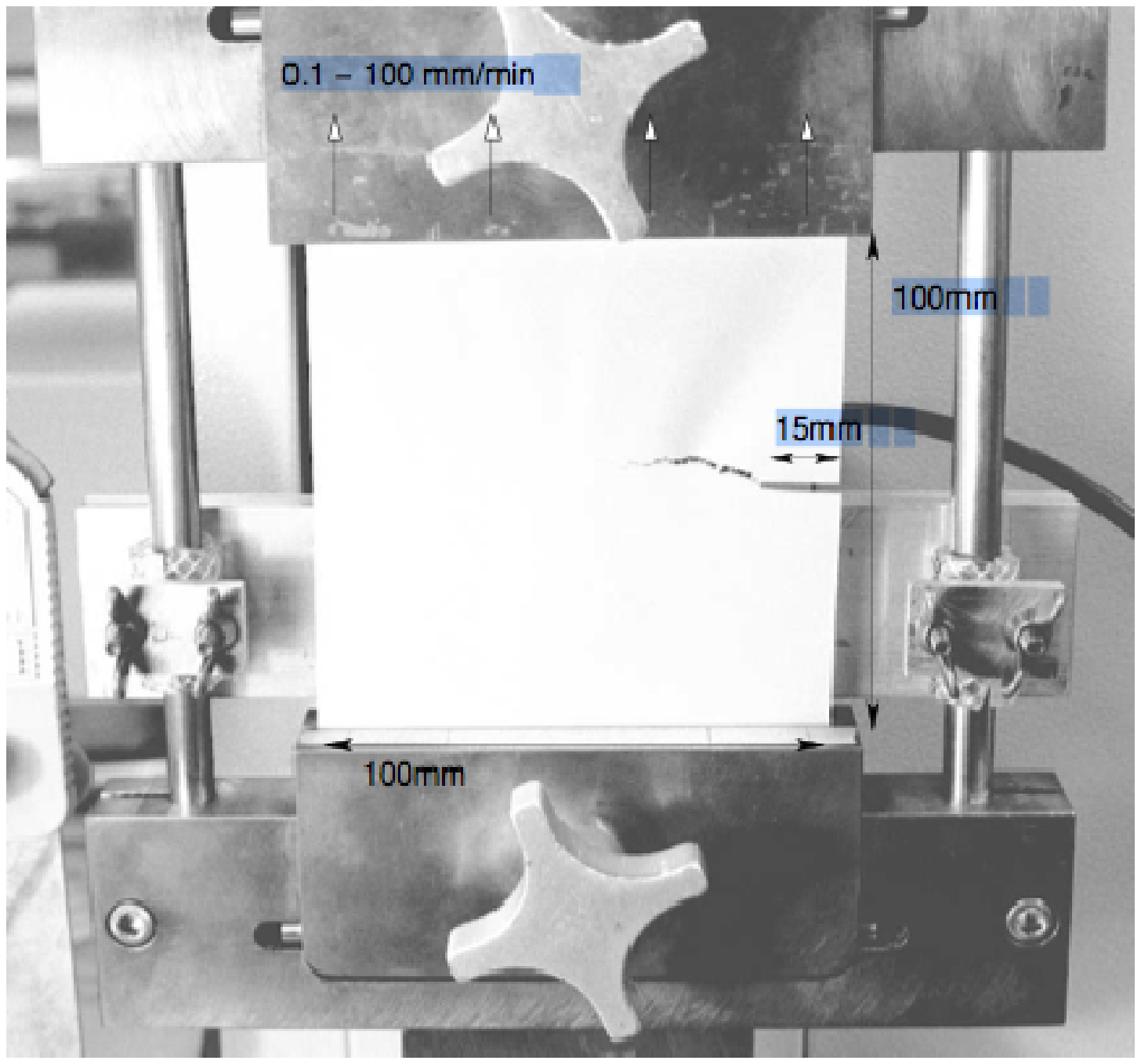,height=7.0cm} \epsfig{figure=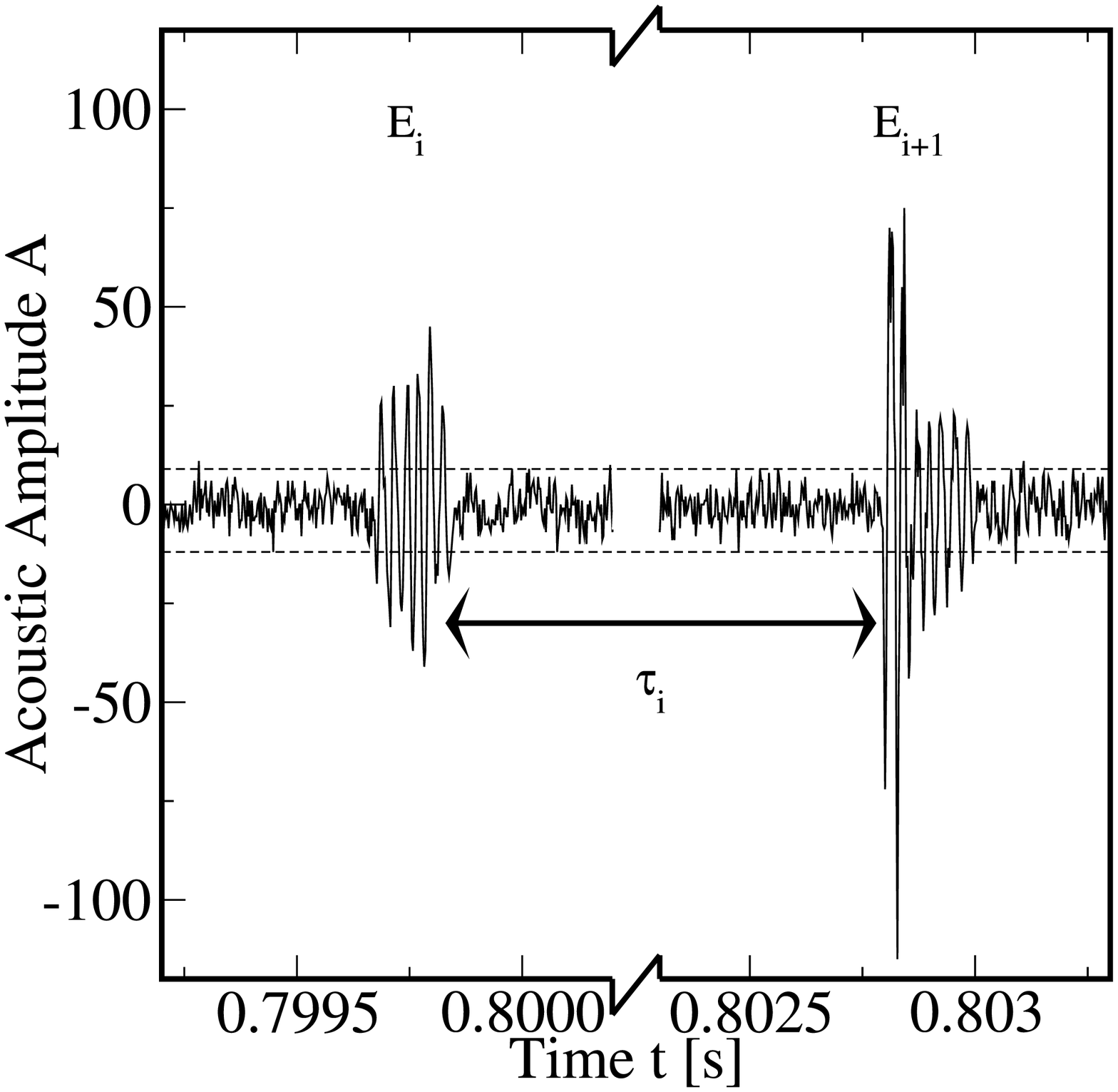,height=5.5cm}

\caption{The measurement setup: sample attached to MTS tensile testing machine. In this Figure single piezocrystal sensor is attached to back of the sample (actual setup consisted two piezocrystals). The figure on the right shows single AE event signal \cite{laurinprl}.}
\label{fig:modes}
\end{figure}

\begin{figure}
\centering
\includegraphics[width=0.8\textwidth]{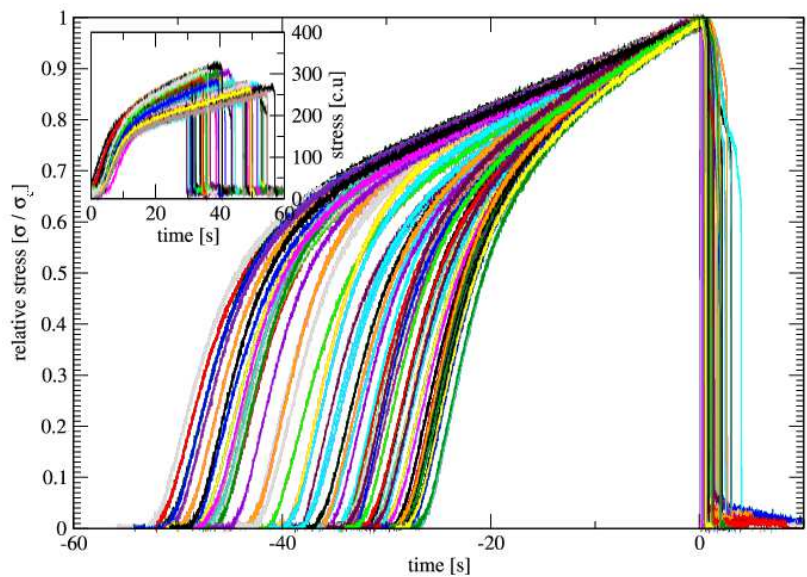}
\caption{Time vs. stress curves from all samples without notch. The maximum
stress is scaled to unity using $\sigma/\sigma_{c}$. This figure depicts typical stress response and its variations as a function of time. The inset contains same data without scaling of the stress-axis.}
\label{fig:strstr}
\end{figure}

\begin{figure}
\centering
\includegraphics[width=0.8\textwidth]{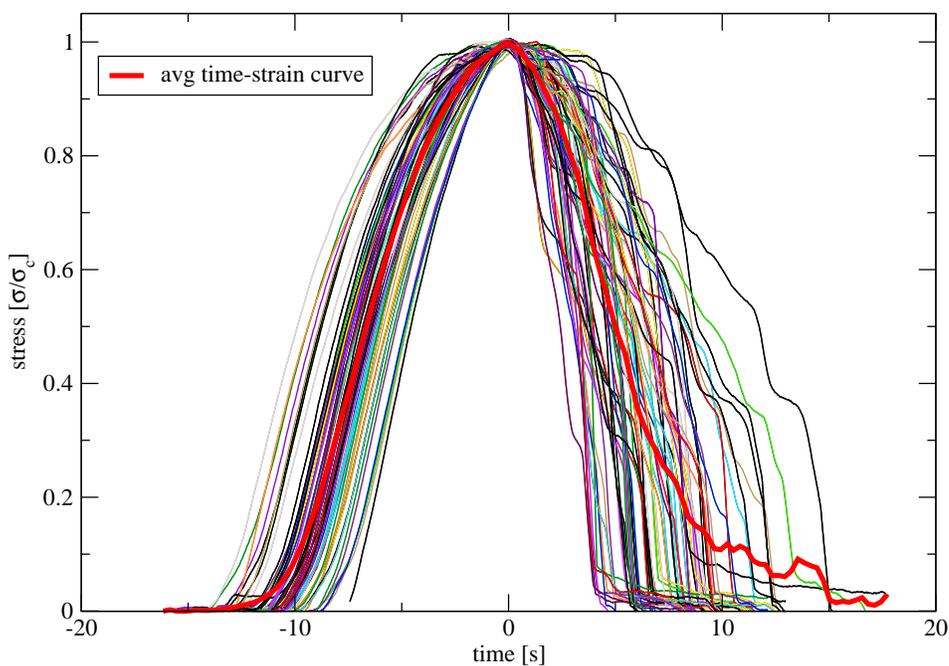}
\caption{Time vs. stress curves from all samples with a notch. The maximum
stress is scaled to unity using $\sigma/\sigma_{c}$. This figure depicts typical stress response of notched samples. Average over all time-stress curves is shown.}
\label{fig:notched-str}
\end{figure}

Figure \ref{fig:modes} shows the measurement setup and a
single AE event. Time vs. stress curves from intact samples are
shown in figure \ref{fig:strstr}, which presents sample-to-sample
variation in different experiments. The curves are shown both after
scaling the sample strength (max. stress) and corresponding time to
unity and in the usual way (inset). Both modes illustrate the
sample-to-sample variation, and also in the proximity of the maximum
stress. Figure \ref{fig:notched-str}  depicts
time-stress-curves with a notch. As it is obvious, $\sigma(t)$ is
of different shape and indicates stable crack growth.

\section{Results}
\subsection{Temporal characteristics of the AE}

We start our analysis from a time ordered set of time-event pairs from an experiment:
$\lbrace (t_1, E_1), (t_2, E_2), ..., (t_n, E_n)\rbrace$. We divide the time interval $(t_1, t_n)$
into windows of length $\Delta t$ and compute the event
energy rate $\dot{E}(t)$ and event count rate $\dot{n}(t)$ for acoustic emission in each
$\Delta t$ window defined according to equations \ref{eq:enrate} and
\ref{eq:countrate}.

\begin{equation}
\dot{E}(t) = \sum_i \frac{H(t-t_i) H(t_i - t + \Delta t) E_i}{\Delta t}
\label{eq:enrate}
\end{equation}

\begin{equation}
\dot{n}(t) = \sum_i \frac{H(t-t_i) H(t_i - t + \Delta t)}{\Delta t}
\label{eq:countrate}
\end{equation}
where $H(x)$ is the Heaviside step function.

After energy and event count rates have been obtained, these
quantities are averaged over all experiments. For the averaging, we 
have to use a similar choice for the time axis (or
stress) and there is no natural choice. The time $t=0$ corresponds to
triggering from the beginning of the tensile experiment, which is unphysical
for any averaging. Before computing the average, we define origin of time and
shift the signals $\dot{n}(t)$ and $\dot{E}(t)$ accordingly.

\begin{figure}[htb]
\begin{center}
\includegraphics[width=1.0\textwidth]{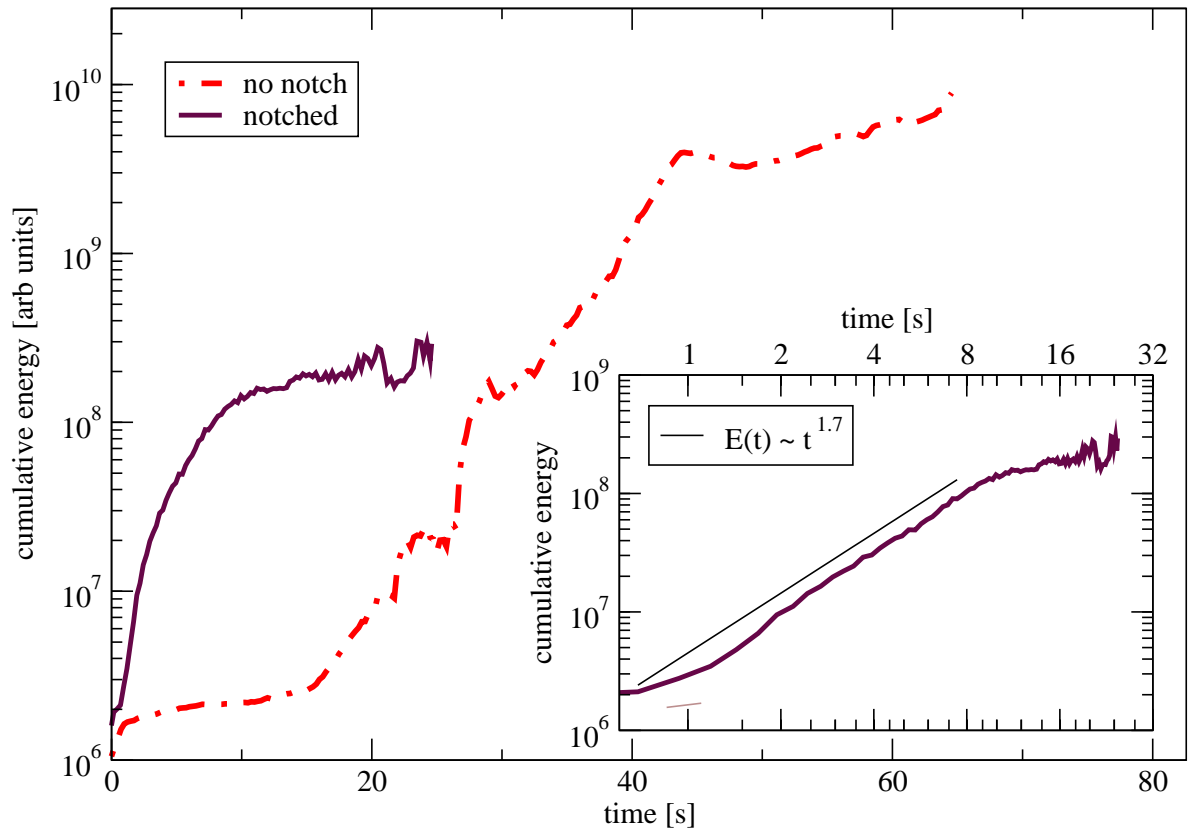}
\end{center}
\caption{Averaged cumulative event energies as a function of time
from the beginning of the experiment. The data is averaged over 100 and 70
samples. The notched case presents a power law increase of acoustic emission energy. The
unnotched highlights the onset of acoustic emission, when the plasticity starts.} \label{fig:aeintegral}
\end{figure}

As a physical beginning of the experiment, the origin of time is defined 
as where we find the maximum slope
of time-stress curve (see figure \ref{fig:strstr}) and the
intersection of the time axis and a straight line fitted to the
point of maximum slope. This choice removes non-idealities at the
beginning of the time-strain curve which often arise to experimental
limitations, e.g. small slack in the sample when it is inserted to
clamps. Figure \ref{fig:aeintegral} shows $E(t)$, the integral of
event energy $\dot{E}(t)$, averaged so that the time is shifted to the beginning of
experiments. The notched case shows stable crack growth.
The unnotched energy integral depicts the onset of the acoustic emission when the 
plastic regime starts, but since time-to-failure scatters from sample to sample,
the meaning of the average becomes subtle. The
energy integral in the unnotched case is similar to the result
presented in \cite{laurinprl}.  The differences between the
result in  \cite{laurinprl} and figure \ref{fig:aeintegral} are
due to the fact that latter does not include events after the stress
has started to drop after maximum has reached, the larger scatter in failure
times and the samples being more ductile. It is interesting to
note that  the data  for the notched samples shows over two orders
of magnitude in integrated energy a power-law scaling $E_{cum}
\propto t^{1.6}$. Assuming stable crack growth with a constant FPZ
size, this feature could be interpreted as the crack length time
dependence. Note that there is roughly one order of magnitude of
difference between the samples without and with notches in  total
energy release magnitude.

\begin{figure}
\centering
\includegraphics[width=1.0\textwidth]{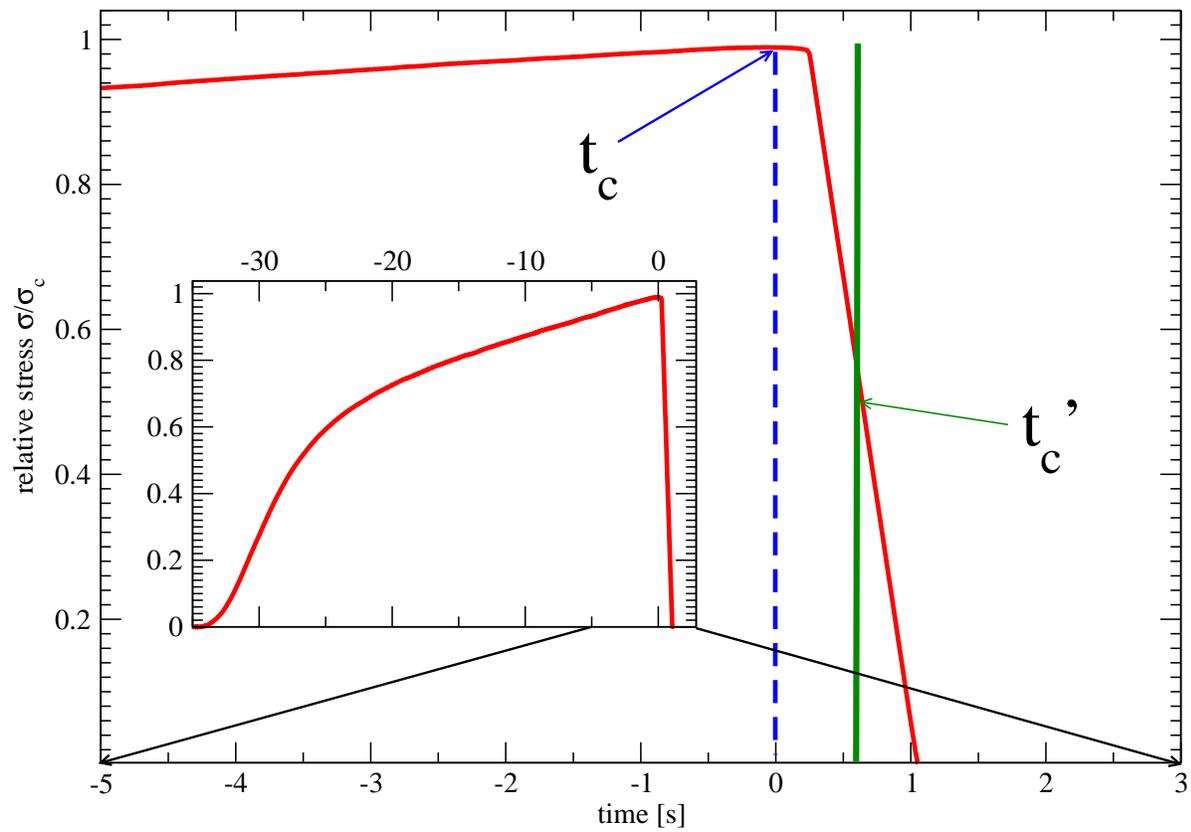}
\caption{Schematic differences of $t_c|\mathrm{max}(\sigma(t))$ and
$t_c'|\mathrm{max}(\dot{n}(t))$ in a
    single experimental time-stress curve. Inset shows time-stress-curve and main panel is magnified to the end
    of the experiment. }
\label{fig:tcschema}
\end{figure}

\begin{figure}
\centering
\includegraphics[width=1.0\textwidth]{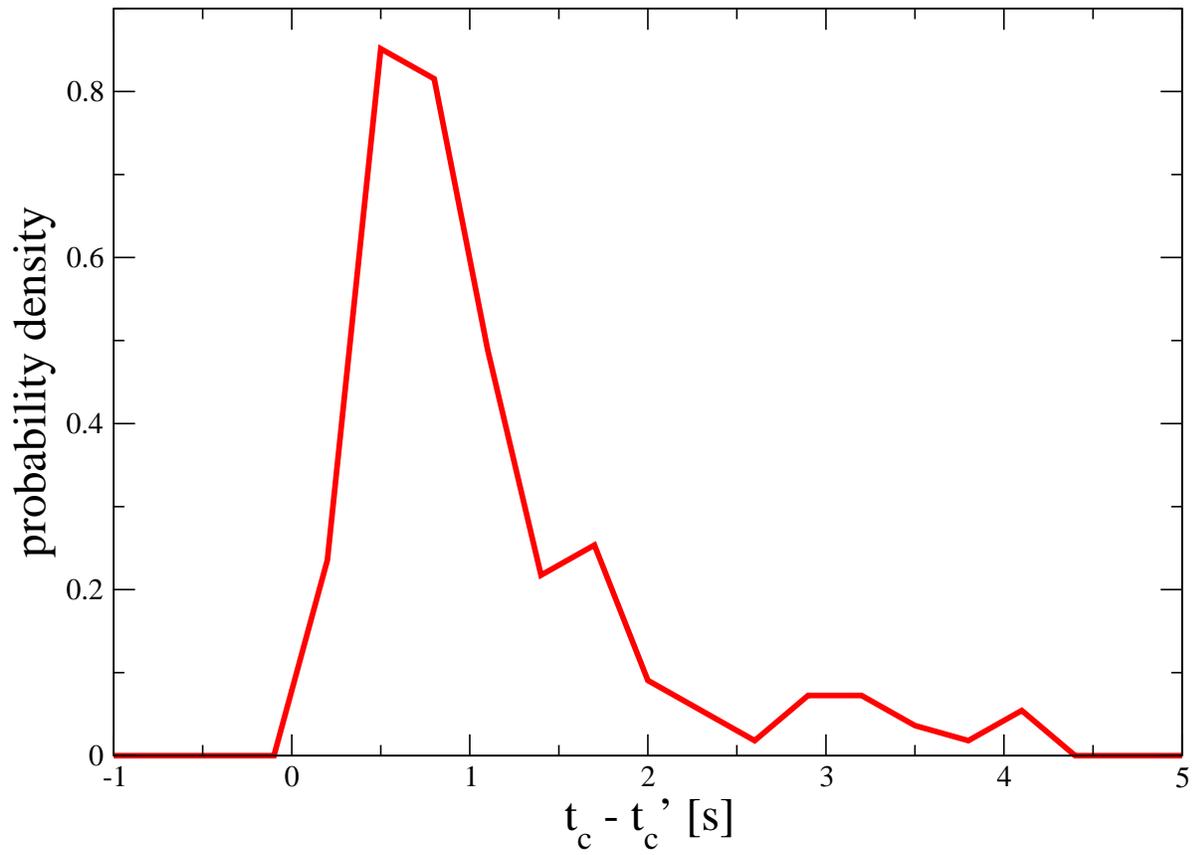}
\caption{The distribution of $t'_c|\mathrm{max}(\dot{n}(t)) -
t_c|\mathrm{max}(\sigma(t))$. The result indicates that time at the maximum
of the event rate is larger than the time at the stress maximum. }
\label{fig:tcafterdistr}
\end{figure}

Next we test for apparent criticality, which would imply the
divergence of the event energy rate near a critical point that is,
the finite lifetime of the sample (corresponding to maximum stress,
or crack instability). Discussing this is meaningful only in the
unnotched case, where one can see a rapid increase in the event
energy near the maximum of the time-stress curve. Conventionally, the
lifetime $t_c$ is defined as the time at the maximum of stress
$\sigma(t)$, that is $t_c|\mathrm{max}(\sigma(t))$. However we
discover that the time at maximum stress $t_c$ is different from the
time at event the  maximum of energy and event count rate
$t_c'|\mathrm{max}(\dot{n}(t))$. Schematically these differences are
shown in the Fig. \ref{fig:tcschema}. The sample to sample
distribution of the quantity $(t_c' - t_c^*)$ is seen in figure
\ref{fig:tcafterdistr} which shows that there is clear difference
time at maximum event rate being larger. In other words, $t_c'$
is found without exception after the maximum stress.

\begin{figure}
\centering
\includegraphics[width=1.0\textwidth]{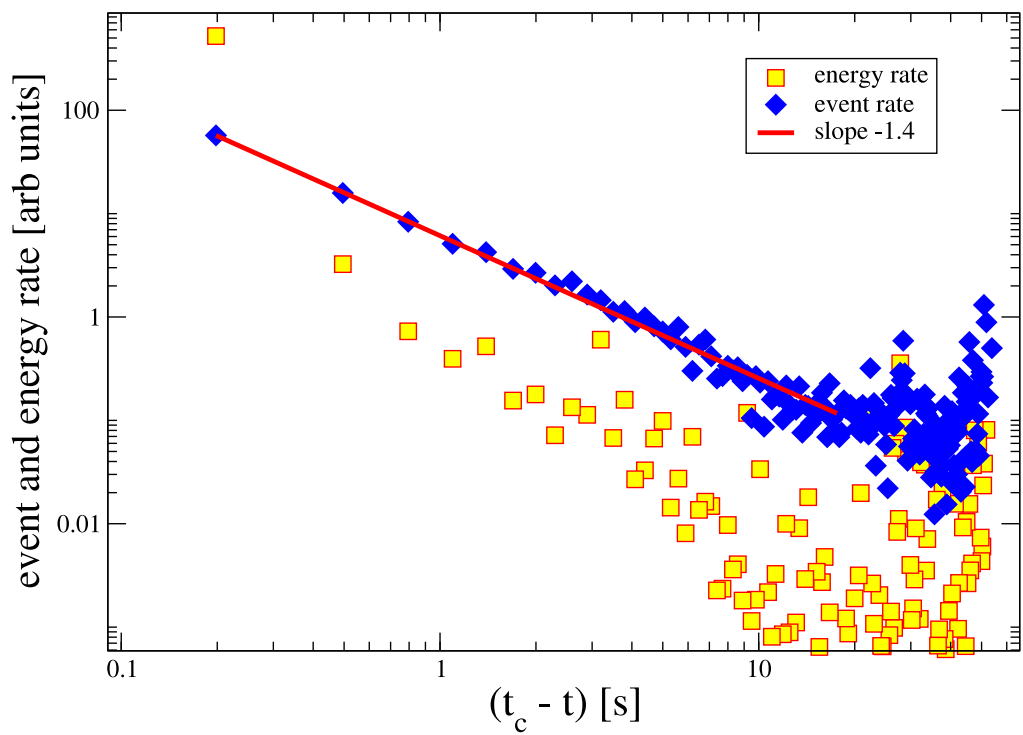}
\caption{The event energy release rate $\dot{E}$ and the event rate $\dot{n}$ as
a function of $t_{c}-t$. Energy release and event rates are computed in a window $\Delta t$=0.2s
which is slided over AE time series. Critical time $t_c$ is defined as a time where
the energy relase rate $\dot{n}$ reaches its maximum value.
Event energy and event rates are averaged over 100 experiments using AE from unnotched samples. The event rates decay as a power law, which is not true for event energies.}
\label{fig:tc}
\end{figure}

When we average quantities over samples by shifting the critical time to
the origin before the averaging, as shown in figure \ref{fig:tc}, we find  that event count rate
$\dot{n}(t_c-t)$ appears to follow a power law according to
Eq.~(\ref{eq:dotnplaw}), with the lifetime as
$t_c|\mathrm{max}(\dot{n}(t))$:

\begin{equation}
\dot{n}(t_c-t)Ê\sim (t_c-t)^{-\Delta } \label{eq:dotnplaw}
\end{equation}
where we find the value $\Delta=1.4\pm0.1$.
The error is defined using the min-max method.

\begin{figure}
\centering
\includegraphics[width=1.0\textwidth]{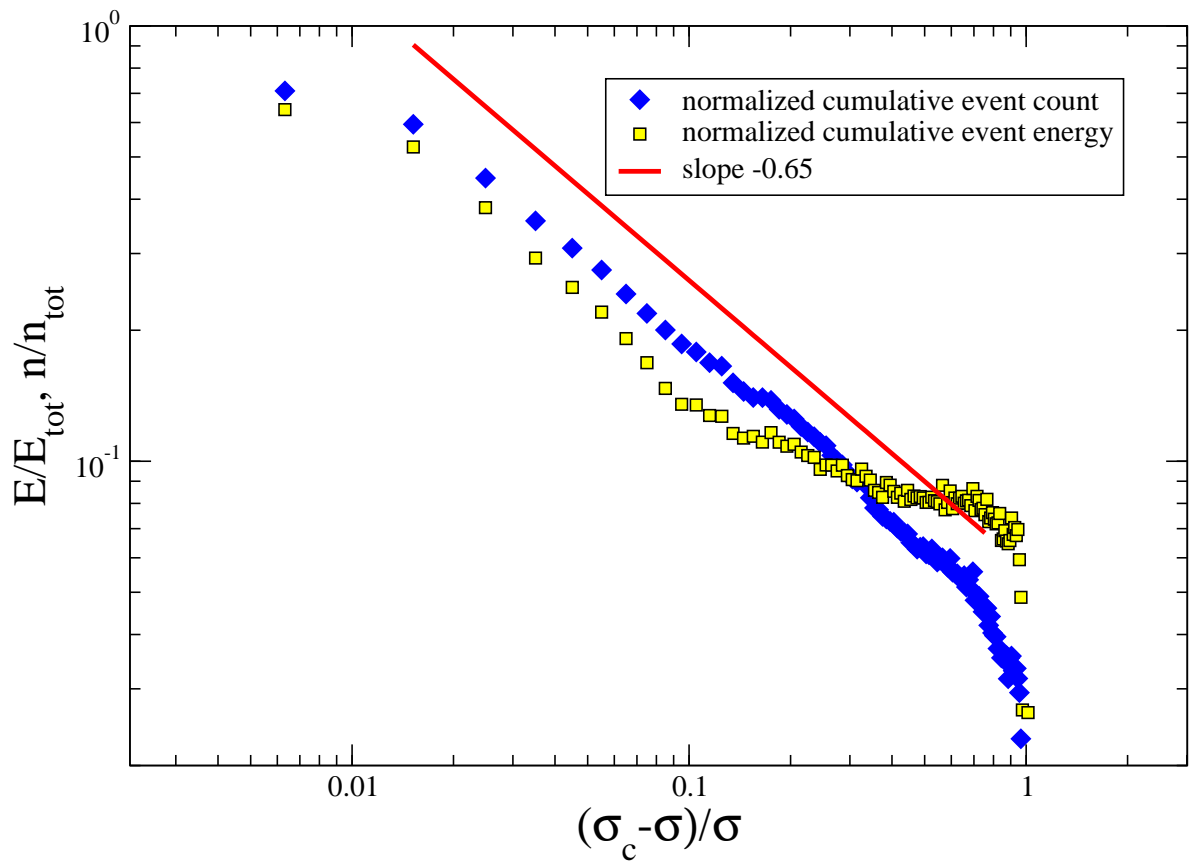}
\caption{Cumulative energy $E/E_{tot}$ and cumulative event count $n/n_{tot}$ as a function of
$\frac{\sigma_c - \sigma}{\sigma_c}$. Event energy and event counts are averaged over 100
experiments using AE from unnotched samples. }
\label{fig:cilib}
\end{figure}

\begin{figure}
\centering
\includegraphics[width=1.0\textwidth]{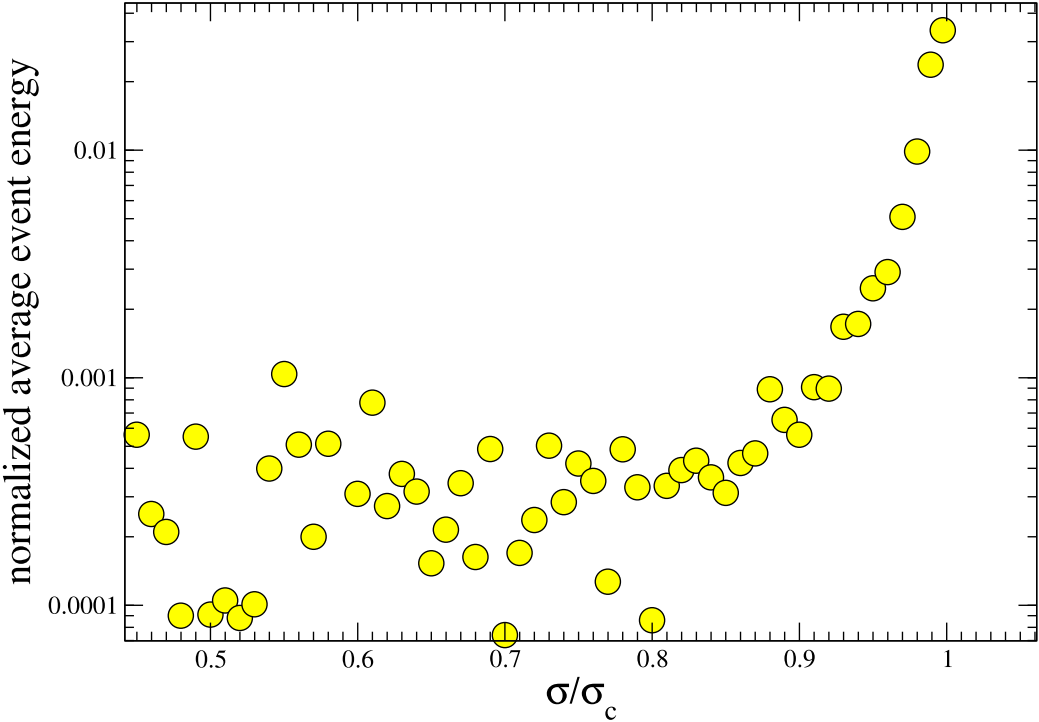}
\caption{Average event energy, $\frac{\dot{E}}{\dot{n}}$, as a function of $\sigma/\sigma_c$. This result
shows that the event energy is not constant during the experiment.}
\label{fig:averageeventenergy}
\end{figure}

Equation (\ref{eq:dotnplaw}) does not apply for the event energy
release rate $\dot{E}(t_c-t)$. A power law behaviour for the event count
rate but not for the event energy implies a change in the average event
energy during the experiment. This is shown in the figure
\ref{fig:averageeventenergy}: average event energy as a
function of normalized stress $\sigma/\sigma_{max}$ during the
experiment. The result shows that there is rapid increase in the
average event energy near $\sigma_c$. We can also look at event and
energy rates as functions of normalized stress $(\sigma_c -
\sigma)/\sigma_c$ as shown in figure
\ref{fig:cilib}. The cumulative event count shows over a
range of rescaled stresses an apparent power-law dependence.
The cumulative event energy behaves in a
way similar to what was found by Garcimartin {\it et~al}
\cite{garcimartinprl}, but a power-law dependence is not clear, when we
have the cumulative count as a reference..
The comparison is not straightforward due to
different loading modes. The connection is further explored in the
Discussion. As expected (on the basis of equation~(\ref{eq:dotnplaw}))
the event rate does not seem to scale here in any particular
fashion.

\subsection{Probability distributions of AE}
Next we study statistical distributions of event energies $E_i$ and
inter-event times. Inter-event times in an experiment are defined as
set of $\tau_i = t_{i+1} - t_i$. These exhibit almost without
exception power-law statistics such that one defines probability
distributions $P(\tau)$ and $P(E)$ for the inter-event/waiting times
and energies which are characterized by power-law exponents $\alpha$
and $\beta$, respectively according to equations \ref{eq:beta} and
\ref{eq:alpha}.

\begin{equation}
P(E) \sim E^{-\beta}
\label{eq:beta}
\end{equation}

\begin{equation}
P(\tau) \sim \tau^{-\alpha}
\label{eq:alpha}
\end{equation}

We find values $\beta = 1.4 \pm 0.1$ and $\alpha = 1.3 \pm 0.2$ for 
event energies and inter-event times respectively.
However, interpretation of these statistical distributions is difficult since
a typical tensile test (or fatigue or creep test) is not stationary.
The internal state of the material changes along the whole test
duration, and statistical quantities such as $P(E)$ or $P(\tau)$
integrate over the whole history of a sample. Thus we study
how these statistical distributions change e.g. in relation to $t_c$
or event count rate $\dot{n}$.

\begin{figure}[htb]
\begin{center}
\includegraphics[width=0.7\textwidth]{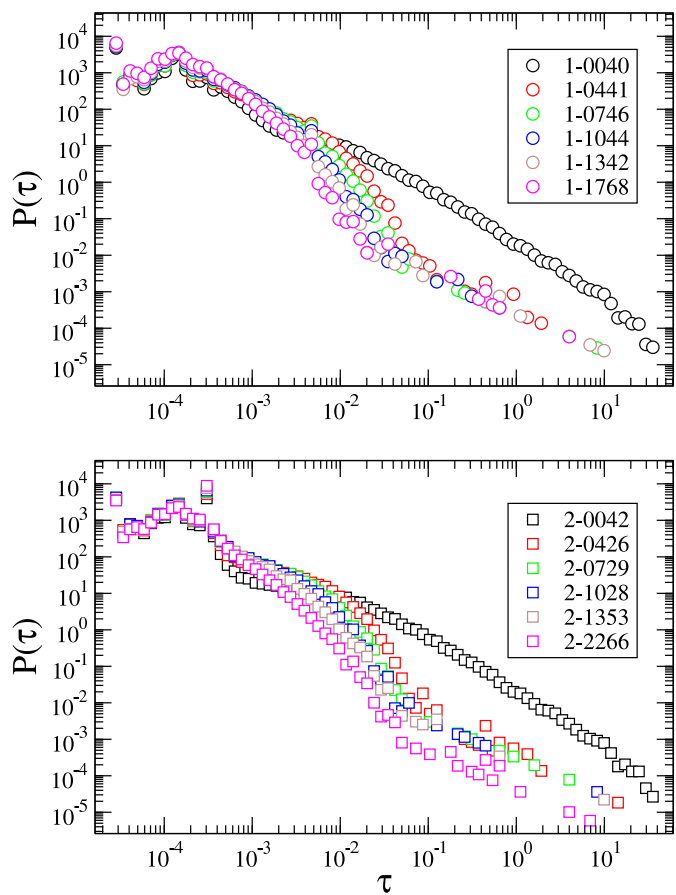}
\end{center}
\caption{Waiting time distributions from tensile
  experiments. AE time series are divided to $\Delta t$=0.2~second time windows and
  waiting times $\tau_i$ and the event rate $\dot{n}$ is computed in a window.
  Windows are divided to different classes based on an event rate
  in the window and the data set in the figure indicates waiting
  time distribution in the event rate class. Label is the averaged event
  rate in the event rate class.  } \label{fig:wt}
\end{figure}

In figure \ref{fig:wt} we show waiting time distributions from tensile
experiments as a function of the event rate $\dot{n}$. The event
rate $\dot{n}$ is computed in a window. Windows are divided into
different classes on the basis of an event rate. The data sets in the
figure represent waiting time distribution in the event rate class.
The data set label indicates the averaged event rate in units
of $s^{-1}$ in the event rate class.  Event rate classes are defined so that
each class contains approximately 20~000 events.  The difference between panels is the algorithm which
identifies events: algorithm used in the lower figure allows overlapping
of AE events while upper does not.  The implication is
that the algorithmic details are not important and differences
we see in distributions is not due to the clustering of events at larger
event rates. From waiting time distributions, it
appears that there is a complicated change of the waiting time
distribution $P(\tau)$ as a function of event rate though one must
be careful as the window size is still (in the time-domain)
relatively large compared to $t_c$. Perhaps, one can see signs of
two different power-laws. It is interesting to note that the purest
power-law behavior is obtained at small event rates (small strains)
while for larger ones it changes to a more complicated distribution.

\begin{figure}
\centering
\includegraphics[width=.7\textwidth]{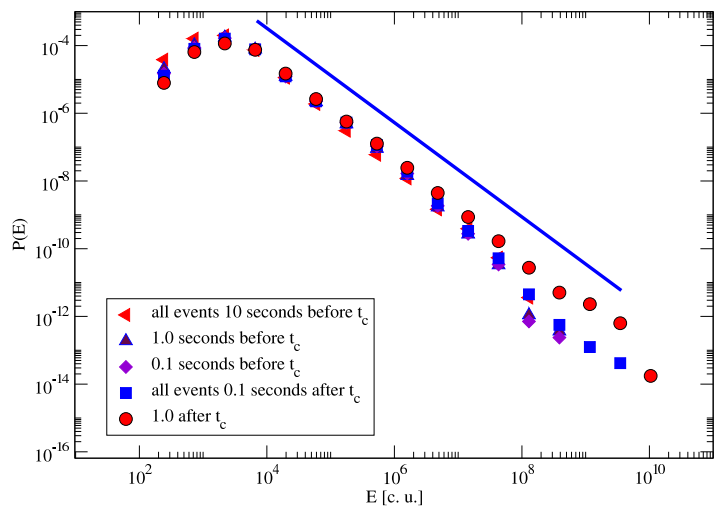}
\caption{The figure depicts evolution in the probability density of event energies when the time at the maximum stress $t_c$ is approached. The distribution contains events which are cumulated up to the time $t_c - t'$. The time shift $t'$ is shown in the label.The distribution becomes broader when the maximum event rate is approached. The solid straight line is a power-law fit $P(E) \sim E^{-1.4}$. }
\label{fig:ratehisttc}
\end{figure}

The energy distributions are shown in figure \ref{fig:ratehisttc}.
Due to large variation in the average event size (see figure
\ref{fig:averageeventenergy}) near time at maximum stress, we studied
changes in the probability density $P(E)$ close to $t_c$. The distribution in the Fig. \ref{fig:ratehisttc} contains events up to the time
$ t_c - t'$. The time shift $t'$  is shown in the label of the data set. The number of events is 6448 if we choose all events up to 10 seconds before $t_c$, and 162425 for all cumulated events up to 1~second after $t_c$. Despite the large increase in
the average event energy, overall form of the probability density function does not change as
much as $P(\tau)$ changes as a function of the event rate. Large event energies are cut off, when one considers the distribution which contains events up to the time which cuts off
events after the event size has started to increase.

\subsection{Theoretical models}

In this chapter we compare experimental results to a random fuse network with a
residual conductivity (RRFN).   The model is analogous to the one
proposed by Duxbury and Li \cite{li}. It is a minimal quasi-static
model for fracture, which includes stress enhancements and residual
bonding, leading to apparent plasticity and crack arrest. The model
is an electrical analogy of an elastic lattice. The main motivation
for using RRFN is that usual RFN simulations are much further away
from the experiments.

In the model we connect a set of fuses to a square lattice with a
size $L \times L$. Fuses are labeled as index $j$ and  associated
with a constant conductivity $g_j = 1.0$ and with randomly distributed
critical currents $i^c_j$. The distribution is uniform and
characterized by its width: $i^c_j \in [1-W, 1+W]$. For the breaking
process we apply a slowly increasing voltage across the lattice
until the hottest fuse approaches its critical current $i^c_j$. The
important ingredient over the usual RFN is that fuses break at two
phases: at the first phase its conductivity drops from $g_j = 1.0$ to a
residual conductivity $g_j = r$.  The fuse remains non-broken in the network
with a residual conductivity until its current threshold is exceeded again and then its conductivity drops to zero, which finally corresponds to breaking the fuse and generating an event. We apply this rule to the
conductivity of the hottest fuse and start increasing voltage from
zero again. The process of applying conductivity drops and
re-evaluating the hottest fuse is repeated until the conductivity of
the network goes to zero and burned fuses form a percolating path
across the lattice.

When the residual conductance approaches unity, the model is
identical to the perfectly brittle random fuse network. When the
residual conductance is close to zero, stress enhancements create
very tough fuses which are capable of arresting cracks and leading
to plasticity. Width of the disorder parameter $W$ leads to
competing effect of the disorder and stress enhancements.

\begin{figure}[htb]
\begin{center}
\includegraphics[width=1.0\textwidth]{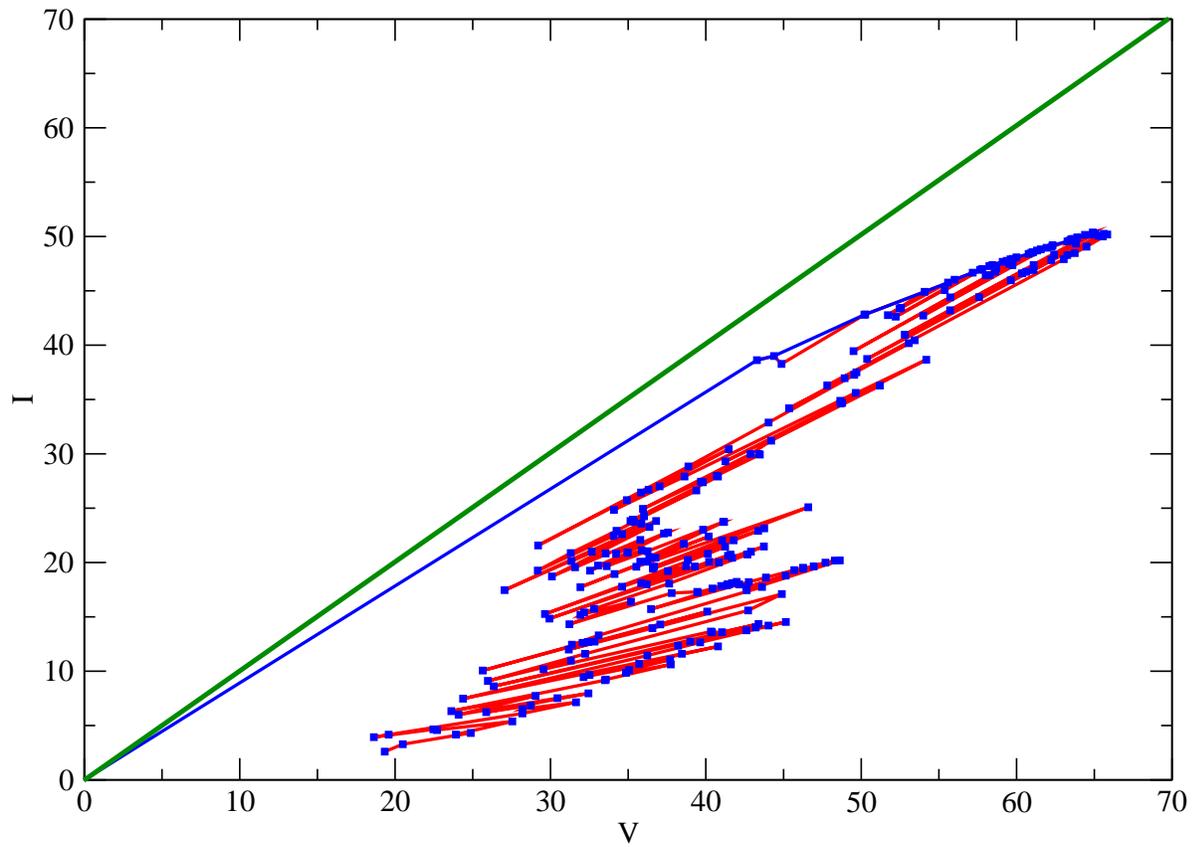}
\end{center}
\caption{Voltage-current pairs from single RRFN simulation. System size is $L =124$, disorder parameter is $W=0.8$ and residual conductivity $r=0.2$. Filled squares indicate VI-pairs when a fuse is burnt, that is its conductivity drops to zero. The straight line corresponds linearly elastic behaviour. The envelope curve corresponds stress-strain curve of the numerical experiment.} \label{fig:visingle}
\end{figure}

\begin{figure}[htb]
\begin{center}
\includegraphics[width=1.0\textwidth]{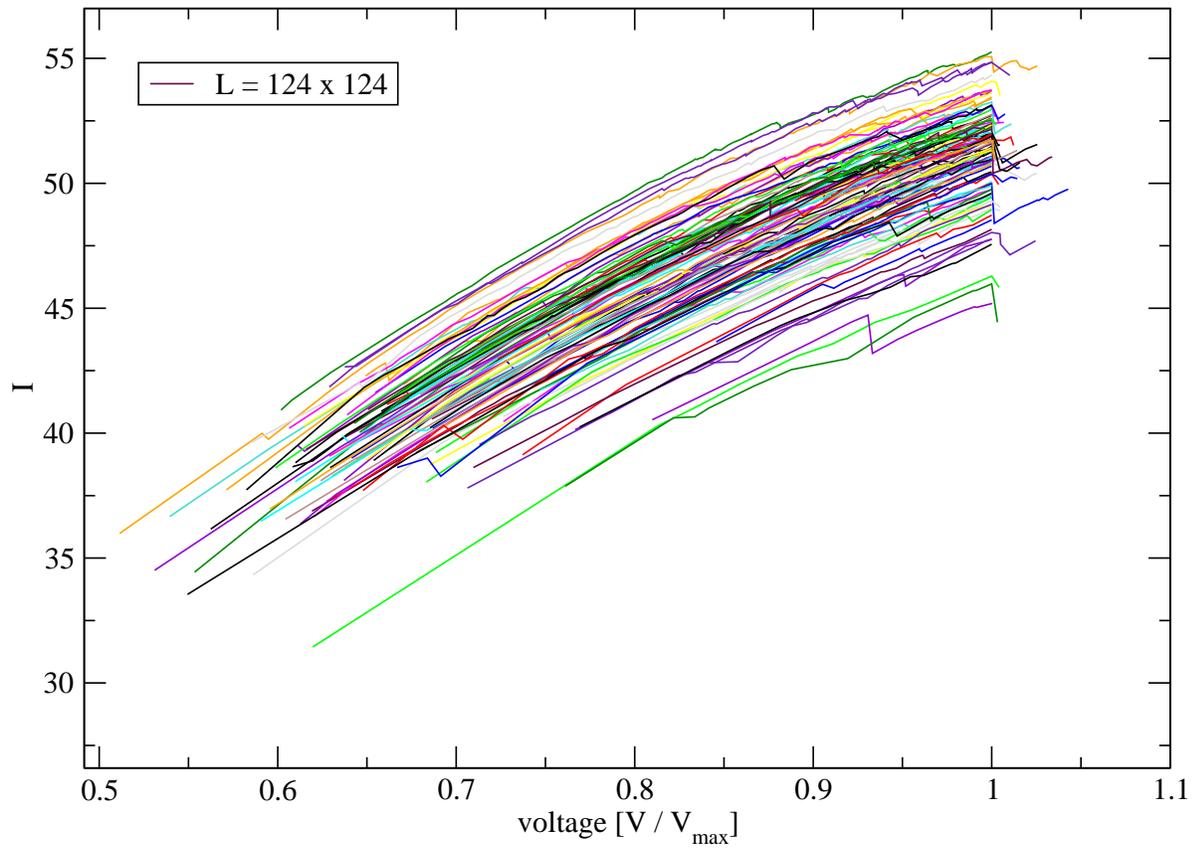}
\end{center}
\caption{99 voltage controlled VI-curves from RRFN simulations with system size $L =124$. The disorder parameter is $W=0.8$ and residual conductivity $r=0.2$. The behaviour correspond qualitatively experimental stress-strain curves.} \label{fig:viall}
\end{figure}

By finding VI-characteristic of the whole network, we obtain an
entity which corresponds experimental stress-strain curves. We
record a  voltage-current pair $(V_i', I_i')$ when a fuses
conductivity drops to zero. We apply voltage-control for the ordered
set $\{(V_i', I_i'),...\}$ by requiring that when the voltage $V_i'$
overcomes the highest value previously recorded we get one point for
the VI-curve $(V_k, I_k)$, where $k$ identifies an "acoustic
emission" event. In the Fig. \ref{fig:visingle} we show all
voltage-current pairs ${V_i', V_i'}$ from one numerical experiment
and the envelope corresponds to the VI-curve $\{(V_k, I_k), ...\}$.
In the Fig. \ref{fig:viall} are voltage-controlled VI-curves from 99
simulations for the system size 124.

The event count rate is defined similarly as in experiments. We
divide voltage-axis  $\Delta V$ sized windows and compute number of
events in a window according to Eq. \ref{eq:simcountrate}. $V_{max}$
is the voltage when  the current is in its maximum. The "acoustic
emission" event in this model is associated to second conductivity
drop of the fuse, i.e. when the conductivity drops to zero. Creating
an event from all conductivity drops was also calculated, but this
did not lead to an acceleration of the event rate.

 %%% meidŠn tavallamme johtavuuden tippuminen pelkŠstŠŠn nollaan luo melua
\begin{equation}
\dot{n}(V/V_{max}) = \frac{\sum_k H(V-V_k) H(V_k - V + \Delta V)  }{\Delta V}
\label{eq:simcountrate}
\end{equation}

We performed set of simulations from system sizes $30 \times 30$ up
to $124 \times 124$. The combination of disorder parameter $W
\approx 0.8$ and  residual conductivity $r \approx 0.2$ led to
similar VI-characteristic as in experiments: activity started after
the experiment approached half of the critical voltage and we
observed activity after the maximum voltage was approached (Fig.
\ref{fig:viall}). This implies that even large cracks are arrested
before a dominating crack is formed.

\begin{figure}[htb]
\begin{center}
\includegraphics[width=1.0\textwidth]{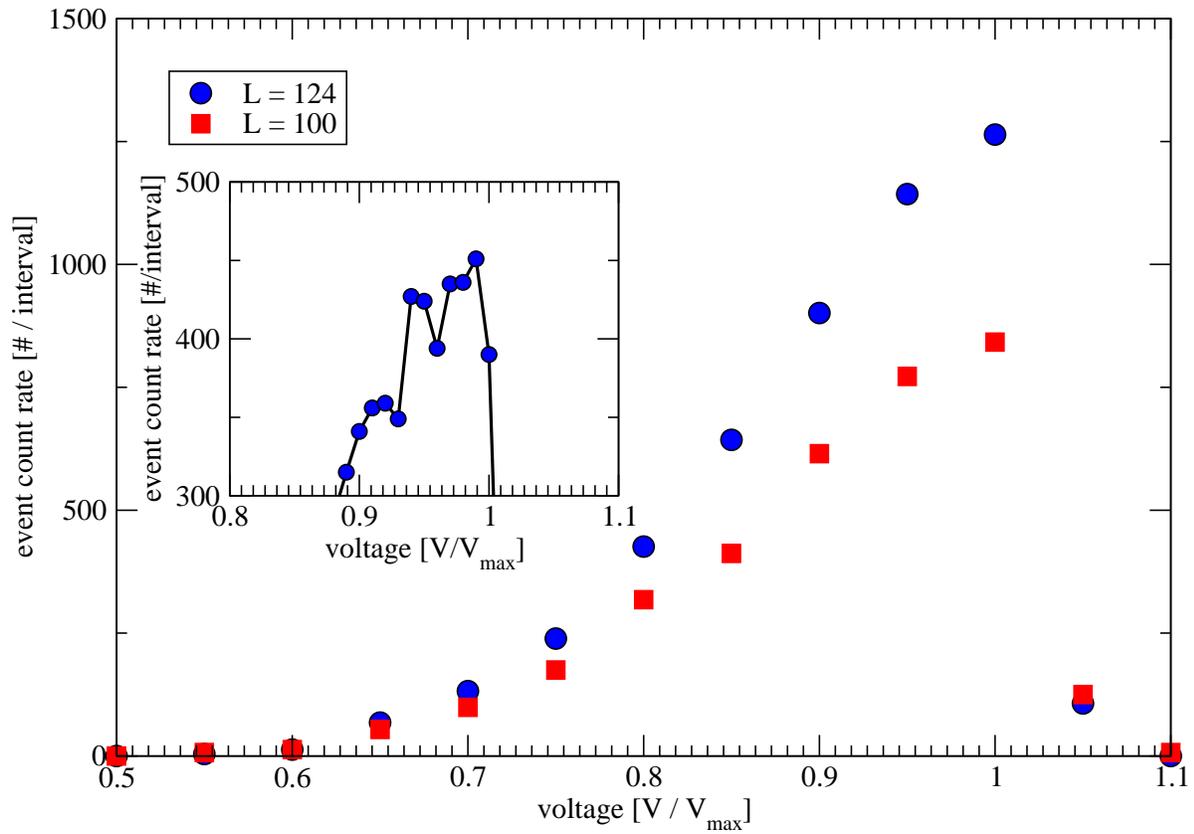}
\end{center}
\caption{Event count rate near critical voltage from RRFN simulations. The data is averaged over 99 samples. Disorder parameter is $W=0.8$ and residual conductivity $r=0.2$. Increase in the event count rate is neither exponential nor a power-law. The tail of the event rate after the maximum current becomes steeper when the system size increases. The inset is a magnifigation of data with ten times smaller bin than in the main figure.} \label{fig:virate}
\end{figure}

\begin{figure}[htb]
\begin{center}
\includegraphics[width=1.0\textwidth]{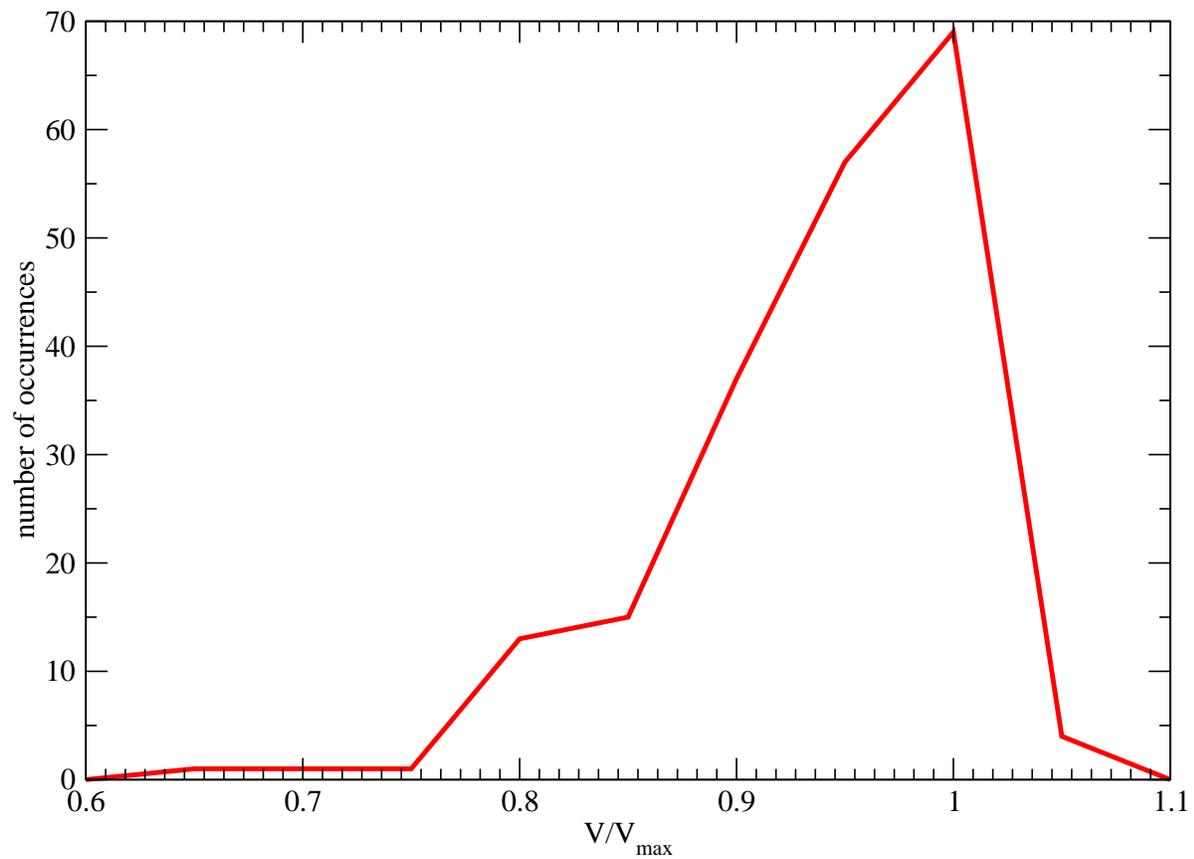}
\end{center}
\caption{Histogram of maximum event rate occurrence as a function of $V / V_{max}$. The result shows sample to sample variation of the relative location of the event rate maximum. Histogram is from $L=124$ and it contains 99 samples. } \label{fig:voltagemaxhistogram}
\end{figure}

 Fig. \ref{fig:virate} shows the event count rate as a function of
$V/V_{max}$ averaged over 99 experiments. The event count rate
starts to increase after 50\% of the maximum voltage is approached.
The event rate is in its maximum near $V_{max}$. There is a small
tail in the event count rate after the maximum stress, which becomes
steeper when the system size increases. The increase in the event
count rate is neither a power law nor critical. In the Fig.
\ref{fig:voltagemaxhistogram} there is the histogram of the maximum
event rate a a function of $V/V_{max}$. This depicts the difference
between experiments and simulations: the maximum event rate occurs
before or at the maximum current in simulations while in experiments
it is observed after the maximum stress. However, there are still a
few occurrences of the maximum event rate after the maximum current
in simulations. We conclude that the time at the maximum event rate
and the time at the maximum current are not identical. However, the
relation is inverse to the experiment.

\section{Discussion}

We have compared fracture precursors using different "critical
points" $t_c | \mathrm{max}(\sigma_c)$ and $t_c'|\mathrm{max}(\dot{n})$ in mode I
loading and  imposing a constant strain rate for copy paper samples.

A divergence is observed in samples without a notch, using time as a
control parameter and looking at the event rate when approaching
time at  the maximum event rate. We have shown that the
characteristic behaviour of the event rate $\dot{E}$ is different
from the event rate $\dot{n}$. The behaviour of the event rate
$\dot{n}(t_{c}-t) \sim t^{-\Delta}$ might be taken to imply
criticality, when $t_{c}$ is chosen to be the time at the maximum of
the event rate. We note that the maximum of the event energy and event
rate corresponds to a time which is observed after the
maximum of the stress $\sigma_c$.

A plausible argument to explain the causal nature of the AE
observations (without a notch) is as follows. At the maximum stress
it follows that $\partial_\epsilon \sigma = E_\epsilon \epsilon + E$
becomes negative. Thus $E_\epsilon$ is negative, which could be
taken to indicate that one of the microcracks dominates and has
started to grow. The event rate increases with the crack growth
velocity, till the crack instability takes place. The divergence of
the event rate could be then arise from a divergence of the crack
velocity. The relation of the rate divergence and the concomitant
increase of the event energy release presents a complicated problem.
The rate of energy release rate should (recall we assume here a
single propagating, stable crack) to be proportional to the new
fracture process zone (FPZ) area created (if the AE event energy is
related to the new FPZ area multiplied by fracture toughness).
However, we have shown that the diverging quantity seems to be the
event rate and not the event energy, in such strain imposed loading.

Earlier experimental studies have shown indications of an event
energy divergence when one imposes a constant pressure rate to a heterogenous material \cite{johansen,
garcimartinprl, nechad}. The critical exponent $\gamma$ in $\dot{E}
\sim \left(\frac{p-p_c}{p_c}\right)^{-\gamma}$ was found to be
$\gamma=1.4$ in \cite{johansen}, $\gamma=1.27$ in
\cite{garcimartinprl} and $\gamma=1.0$ in \cite{nechad}. In the case
of an imposed constant strain rate there was not found to be any
critical divergence of the energy release rate \cite{guariano}. The
analysis however was not done in the same spirit, and excluded the
mechanism suggested above. Finally it was suggested that the real
control parameter is time, since both imposing constant or cyclic
stress seems to imply a critical divergence of the event energy
\cite{guariano2}. Our results include a somewhat similar behavior
for the event energies as the maximum stress is approached.
%Our experimental findings shows
%similar form for the event energy rate, when we apply transformation
%from time to stress (see Fig. \ref{fig:cilibertoenergy}).

Statistical distributions of event energies $P(E)$ and inter-event
times  $P(\tau)$ have been measured earlier several times from
paper. Power laws have been found with exponents $\beta=1.4\pm0.1$
and $\alpha=1.3\pm0.2$ for event energies and inter-event times,
respectively. We may contrast the current result to earlier work by
Salminen et al. where acoustic emission has been measured from paper
in the machine direction \cite{laurinprl}. The energy exponent
$\beta$ is here slightly larger, compared to $\beta=1.25\pm0.10$, 
which might be attributed to the
much more ductile nature of paper as a material when stressed in the
cross-machine direction. The waiting time exponent $\alpha$ is
almost identical when it is integrated over whole experiment.

In any case, in general most of the statistical signals that have
been explored experimentally are still awaiting for theoretical
explanations. For e.g.  $P(E)$, brittle fracture models allow to
derive power-law -like scaling forms but with exponents $\beta$ that
are in general far too large. To see if qualitative agreement could
be found we compared the event rate from experimental data  to the
residual random fuse network. The event rate was found to increase
when the voltage at maximum current was approached - contrary to
perfectly brittle ordinary RFN. The result was not power-law
divergence as in the experimental data. Note that introducing
fracture toughness (residual fuses) to the model caused a change to
the behaviour of the form of the event count rate: rapid increase of
the event energy at $V_{max}$ was not found in the perfectly brittle
RFN model. Further studies of the RRFN would seem to be of interest.
Note that Amitrano and Helmstetter \cite{amitrano} have proposed a
numerical model in order to model time-dependent damage and
deformation of rocks under creep. In the model the 2D finite element
method is used and separate time-dependent and time-independent
damage progression and time-to-failure laws are introduced.
Time-independent damage progression on an element is introduced as
a gradual drop of an elastic modulus if the stress threshold was
exceeded: $E_i(n+1) = E_i(n)(1-D_0)$ where $E_i(n)$ is elastic
modulus of an element after $n$ damage events and $D_0$ is a
constant damage parameter. Quasistatic stress redistribution may
induce an avalanche of damage events during a single loading step.
The damage progression law and quasistatic stress redistribution is
similar to conductivity drop and re-evaluation of the voltage in
RRFN. The difference is in the damage law of an element. The result
of the model is a power-law acceleration of the strain rate
$\dot{\epsilon}$ and the rate of damage events $\dot{n}$ near the
failure time $t_c$. However, if we look at number of events
as a function of strain $\frac{\dot{n}}{\dot{\epsilon}} \sim
\frac{1}{(t_c-t)^{\gamma*} }$ the rate of damaged events decreases
when $t_c$ is approached. The result is qualitatively in agreement
what is observed on the residual random fuse network when it is
assumed that every conductivity drop creates an acoustic emission
event. The gradual failure does not explain the event rate
acceleration near the failure in the experiments.

Eq.~(\ref{eq:dotnplaw}) is in essence identical to the Omori's law
of aftershock rates. In the stationary paper peeling experiment one
observes Omori's law with exponent close to $\Delta=1.4$ and it is
also comparable e.g. to tectonic seismicity. In that case the
generic temporal properties are further discussed in Ref.
\cite{rosti}. Approaching  a``main shock" in stationary experiment
might be identical to approaching time at the maximum event rate in
strain-driven non-stationary fracture experiment when we consider
the behaviour of the event rate $\dot{n}$. Fundamental differences
of non-stationary and stationary AE signals are the different forms
of the inter-event time distribution and the increase in the average
event energy before failure (or main shock). However, this
acceleration of the event rate in the stationary experiment is
significant when averaging over a large number of sequences and the
similarity may be coincidental.
%%Curiously the event count rate  does not
%%distinguish a non-stationary and a stationary signal: activity increases... blaablaa
%%{\bf en tajua viimeista lausetta}

The limitation of the experimental setup is the limited choice
of loading modes and lack of feedback control of the imposed strain or the stress.
For example, driving material based on the
acoustic emission event rate, as applied for the rock fracture in Ref. \cite{davidsen},
we could obtain an experimental test to
the criticality and diverging quantities of the paper fracture; it would
be interesting to see if it is possible to alter the event rate divergence near $t_c$
by controlling the strain based on AE feedback during an experiment.

 {\it Acknowledgements -} The authors would like to
thank for the support of the Center of Excellence -program of the
Academy of Finland, and the financial support of the European
Commissions NEST Pathfinder programme TRIGS under contract
NEST-2005-PATH-COM-043386. Discussions and/or collaborations with
Amandine Miksic and Lauri Salminen are also acknowledged.

\FloatBarrier

%\section{Figures}

\end{document}